\documentclass[a4paper,fleqn]{cas-sc}

\usepackage[numbers,sort&compress]{natbib}
\usepackage{subfig}
\usepackage{ulem}

\def\tsc#1{\csdef{#1}{\textsc{\lowercase{#1}}\xspace}}
\tsc{WGM}
\tsc{QE}
\tsc{EP}
\tsc{PMS}
\tsc{BEC}
\tsc{DE}

\begin{document}
\let\WriteBookmarks\relax
\def\floatpagepagefraction{1}
\def\textpagefraction{.001}

\shorttitle{Payoff-Driven Coevolution in Hypergraph}

\shortauthors{Yichao Yao et~al.}

\title [mode = title]{Payoff-Driven Coevolution and Oscillatory Dynamics in Hypergraph}                      


%
\author[1]{Yichao Yao}
\fnmark[1]

\affiliation[1]{organization={College of Artificial Intelligence, Southwest University},
	city={Chongqing},
	postcode={400715}, 
	country={PR China}}

\author[1]{Yuji Zhang}
\fnmark[1]

\author[2]{Juan Wu}

\affiliation[2]{organization={School of Statistics and Data Science, Xi’an University of Finance and Economics},
	city={Xi’an},
	postcode={710100}, 
	country={PR China}}

\author[1]{Minyu Feng}[orcid=0000-0001-6772-3017]
\cormark[1]
\ead{myfeng@swu.edu.cn}

\cortext[cor1]{Corresponding author}

\author[3]{Attila Szolnoki}

\affiliation[3]{organization={Institute of Technical Physics and Materials Science, Centre for Energy Research},
	city={Budapest},
	postcode={H-1525}, 
	country={Hungary}}

\fntext[1]{These authors contributed equally to this work.}

\begin{abstract}
We study a coevolutionary public goods game on a dynamic hypergraph, where an individual's payoff directly determines the number of hyperedges it can join. In the proposed mechanism, nodes adjust their participation according to the group payoffs of hyperedges, and hyperedges that remain occupied only by defectors for a sufficiently long time collapse and are rebuilt by selecting new members based on the current payoffs of nodes. This adaptive rule captures the performance-driven reorganization of group interactions in evolving collective systems. Using Monte Carlo simulations, we show that the cooperation fraction and average hyperdegree may converge to steady states with stochastic fluctuations or exhibit persistent oscillations, depending on the parameter regime. The steady-state outcomes are strongly nonmonotonic with respect to the structural adaptation parameters: cooperation is sustained only when the rate of link formation is properly balanced. If structural adaptation is too fast, frequent contacts between cooperators and defectors destroy cooperative clusters; if it is too slow, cooperators lack sufficient structural support to expand. This differs from the conventional expectation in static settings that larger benefit parameters always facilitate cooperation. We further introduce spectral entropy to quantify the regularity of the oscillatory dynamics and identify limit-cycle behavior in the phase space in certain regimes. These results suggest that adaptive higher-order restructuring can both promote and destabilize cooperation, offering insight into oscillatory cooperation and recurrent prosperity-decline cycles in real group-structured systems.

\end{abstract}


\begin{highlights}
\item A dynamic hypergraph public goods game with payoff-driven participation is proposed.
\item Cooperation and network density show a nonlinear relation under adaptive dynamics.
\item Spectral entropy quantifies oscillatory regularity and detects limit cycles.
\item The model reveals recurrent cooperation-growth and collapse cycles in systems.
\end{highlights}


\maketitle

\section{Introduction}
Cooperation is a pervasive phenomenon in nature, observable in a wide range of systems, from human societies to biological populations~\cite{nowak2011supercooperators,rand2013human,han2025cooperation}. As a cornerstone of social prosperity and development, the evolution of cooperation has attracted extensive research over the past decades~\cite{traulsen2023future,perc2017statistical}. As a milestone in early research, Nowak identified five fundamental mechanisms that support the evolution of cooperative behavior \cite{nowak2006five}, which were studied in various models such as the Prisoner's Dilemma, the Snowdrift Game, and the Public Goods Game~\cite{nowak1993strategy,doebeli2005models}. In this context, network science provides a natural framework for studying cooperation and offers valuable insights into the emergence of cooperative behavior in real-world systems~\cite{szabo2007evolutionary}. In particular, cooperation on lattices~\cite{nowak_n92b,szabo_pre05}, small-world networks~\cite{masuda2003spatial,kim2002dynamic}, and scale-free networks~\cite{santos2005scale,gomez2008natural} has been widely studied, and numerous evolutionary models and mechanisms inspired by real scenarios have been proposed, including pairwise learning, birth-death processes, reputation systems, and three-player game mechanisms, among others~\cite{luo_c_amm26,feng2024evolutionary,szolnoki2016leaders,feng2023harmful,han_j_csf25,kang_hw_pla24,benko2025evolutionary,zhang_yj_ieee25,pi2024memory}.

In the early stage of research, the evolution of cooperation on networks has been primarily studied on static structures. However, in reality, social environments are dynamic and the interactions between individuals often change over time~\cite{zimmermann2001cooperation}. Individuals’ strategic behavior can, in turn, reshape their social ties, and personal tendencies can influence the overall structure of interactions~\cite{szolnoki2017environmental,jusup2022social}. To capture such feedback, coevolutionary network models have been developed in which both strategies and network structures evolve simultaneously~\cite{perc2010coevolutionary,rand2011dynamic}. Such models include coevolution via random edge rewiring~\cite{pacheco2006coevolution,pacheco2006active}, intermittent activation and inhibition of edges~\cite{zeng2025complex}, and models driven by individual social willingness~\cite{szolnoki2009resolving,yao2023inhibition,zeng2022spatial}, revealing rich and complex evolutionary patterns.

Nevertheless, most of these coevolutionary models focus on pairwise interactions. For group-based games, such as the Public Goods Game (PGG), traditional network formulations often assume that each individual serves as the focal player of a PGG group with all its neighbors~\cite{milinski2002reputation,santos2008social}. Based on this assumption, a wide range of cooperation promoting mechanisms have been extensively investigated, including incentive schemes such as reward and punishment~\cite{szolnoki2010reward,helbing2010defector}, heterogeneity in both individual attributes and network structure~\cite{perc2008social,santos2008social}, as well as reputation and memory based mechanisms~\cite{siyu2025reputation,li2013one}. However, in reality, not every individual necessarily acts as a focal player, and the number of PGG groups in the network does not always correspond one-to-one with the number of nodes~\cite{perc2013evolutionary}. Hypergraphs naturally address this limitation~\cite{alvarez2021evolutionary}. Unlike traditional networks where edges connect only two nodes, hypergraphs allow each edge —— referred to as a hyperedge —— to encompass multiple nodes, thereby capturing group-level interactions~\cite{battiston2021physics,alvarez2021evolutionary}. \textcolor{black}{Recent studies have demonstrated that such higher-order interaction structures can play a crucial role in shaping collective behavior and the emergence of cooperation. For example, it has been shown that higher-order interactions significantly influence collective human dynamics and behavioral patterns~\cite{battiston2025higher}. In the context of evolutionary game theory, a growing body of work has investigated cooperation mechanisms on hypergraphs, including the interplay between reinforcement learning and cooperative dynamics~\cite{xu2025reinforcement,xu2024reinforcement}, as well as the effects of mixed game interactions on higher-order structures~\cite{wang2024mixing}. These studies highlight that incorporating higher-order interactions can fundamentally alter the evolutionary pathways of cooperation, making hypergraph-based models a natural and necessary framework for studying group-based social dilemmas.}

On hypergraphs, PGGs can be defined over hyperedges, where all nodes belonging to the same hyperedge participate in a common public goods game~\cite{civilini2021evolutionary}. Inspired by this perspective, recent studies have explored how hypergraphs facilitate cooperation. For instance, Alvarez-Rodriguez {\it et al.} discussed the game of common goods on uniform hypergraphs and heterogeneous random hypergraphs, explicitly elevating agent interaction to the situation of non-binary interaction~\cite{alvarez2021evolutionary}, Pan {\it et al.} introduced heterogeneous investment into the public goods game on a uniform hypergraph and found its promoting effect on cooperation~\cite{pan2023heterogeneous}, Civilini {\it et al.} investigated the evolutionary game model of group choice dilemmas on hypergraphs, explaining the emergence of irrational herd behavior and radical behavior in social groups~\cite{civilini2021evolutionary,civilini2024explosive}, and Zhang {\it et al.} studied the different influences of hypergraphs and simplicial complexes on the evolution of cooperation~\cite{zhang2023higher}.

To date, most studies of games on higher-order networks still assume static structures, neglecting the feedback of individuals’ strategic behavior on network evolution. In reality, the willingness and intensity of individuals’ participation in group interactions often depend on their accumulated social resources. For example, large corporations with substantial capital tend to engage in more collaborative projects; wealthy investors often diversify across multiple stocks or funds; renowned researchers frequently collaborate with many colleagues. Conversely, small firms may participate in only a few minor projects, modest investors handle fewer assets, and early-career researchers typically collaborate with only a limited number of peers. Motivated by these observations, this work proposes a coevolutionary model of PGG on hypergraphs, in which the number of groups in which an individual can participate is adjusted based on its current payoff. Consequently, highly successful individuals can engage in more PGG groups, while less successful individuals are compelled to leave certain groups. During the process of joining and leaving groups, individuals preferentially join higher-payoff groups and exit groups yielding lower payoffs. This design captures an intuitive aspect of real systems: investors and companies typically prioritize joining more profitable projects while abandoning less lucrative ones.

Motivated by these real-life observations, we now introduce a collapse mechanism for hyperedges, whereby hyperedges with no cooperative participation over extended periods naturally vanish and new hyperedges emerge elsewhere in the network to replace them. This mechanism mimics real-world public projects that fail due to lack of participation or organization. Together, these mechanisms enable the network to adapt dynamically: connections among individuals can increase or decrease, and hyperedges are continuously updated, allowing the coevolution of strategic behavior and network structure to generate rich evolutionary dynamics. In this study, we primarily investigate how the synergy factor of the PGG and the sensitivity of group participation to individual payoff jointly influence the overall cooperation fraction and average hyperdegree in the network.

The remainder of the paper is organized as follows. In Section~2, we describe our model in detail. Section~3 presents the results of evolutionary simulations. Finally, Section~4 concludes with a summary and discussion of some future research directions.

\section{Model}

In this section, we present the details of the model. We first describe the public goods game on hypergraphs, followed by the mechanisms governing the evolution of the network structure. Next, we introduce the strategy update rules for individual agents. Finally, we provide an overview of the overall evolutionary procedure.

\begin{figure}[]
	\centering
	\subfloat{
		\includegraphics[scale=.5]{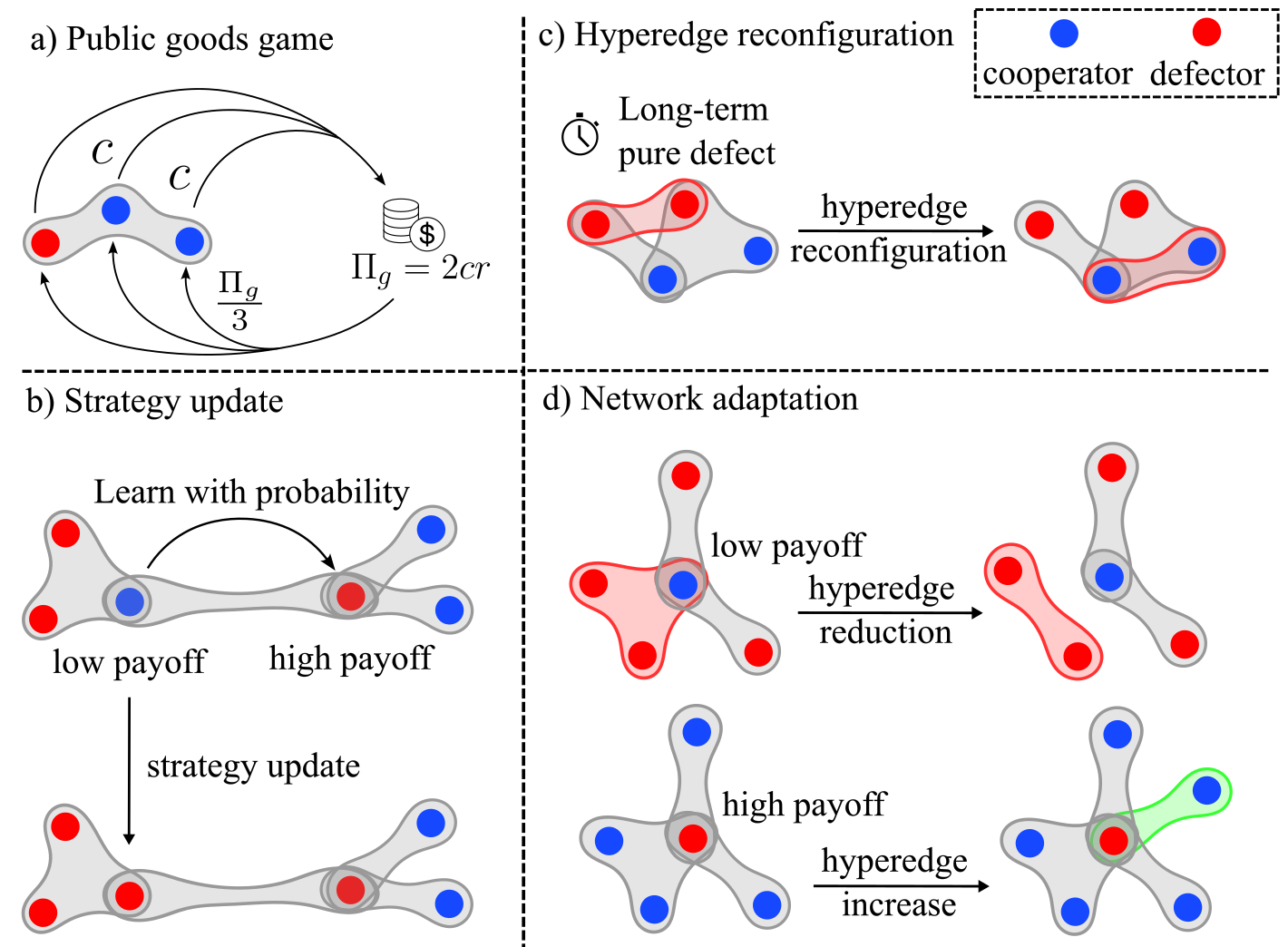}}
	\caption{\label{FIG:0}\textbf{Schematic illustration of the coevolutionary dynamics.} Red and blue nodes represent defectors and cooperators, respectively. 
		a) Illustration of the public goods game, where in a group of three individuals, two cooperators contribute and all members receive the resulting payoff. 
		b) Illustration of strategy updating, where a low-payoff cooperator adopts the strategy of a higher-payoff defector according to the update rule. 
		c) Illustration of hyperedge reorganization, where a red hyperedge, occupied only by defectors for a prolonged period, collapses and is reassigned to other nodes. 
		d) Illustration of structural evolution, where a low-payoff cooperator leaves a red hyperedge that exploits it most severely, while a high-payoff defector joins a new green hyperedge.}
\end{figure}

\subsection{Public goods game on hypergraph}

We first introduce the Public Goods Game (PGG) on hypergraph. In hypergraph, each hyperedge represents a public goods group. At each discrete time step, every individual in the network collects payoff from all the public goods groups it participates in. The joint effort of the group members is represented by a synergy factor $r$. According to the main strategies, cooperators contribute a cost $c$ to each group in which they participate, whereas defectors do not incur any cost.

The payoff of individuals on a public goods group $g$ is calculated as
\begin{equation}
    \Pi_g=\frac{r\Sigma_{i\in g}s_ic}{|g|}\,,
\end{equation}
where $s_i=1$ if individual $i$ is a cooperator and $s_i=0$ if $i$ is a defector, and $|g|$ marks the size of group $g$. The total payoff of an individual $i$ is then
\begin{equation}
    \Pi_i=\Sigma_{g\in G_i}\Pi_g-s_ic|G_i|\,,
\end{equation}
where $G_i$ denotes the set of hyperedges (public goods groups) that individual $i$ participates in. The first term represents the share of the total group benefits received by individual $i$, and the second term represents the total investment made by $i$ if it is a cooperator. An example of a public goods game occurring on a single hyperedge is illustrated in part a) of Fig.~\ref{FIG:0}.

\subsection{Strategy evolution}

After updating the payoff and network structure, individuals update their strategies by imitating a more successful partner based on pairwise comparison of payoff values. At each time step, an individual $i$ randomly selects one of its neighbors $j$ (i.e., an individual sharing at least one common hyperedge with $i$) and compares their payoffs. Individual $i$ adopts the strategy of $j$ with probability
\begin{equation}
	W_{i\leftarrow j}=\frac{1}{1+\exp[-\beta(\Pi_j-\Pi_i)]}\,,
\end{equation}
where $\Pi_j$ and $\Pi_i$ denote the payoffs of individuals $i$ and $j$, respectively, and $\beta$ represents the intensity of selection. Accordingly, a larger value of $\beta$ implies that the payoff difference plays a dominant role in strategy update, whereas small $\beta$ value corresponds to a more stochastic decision-making process.

This update rule allows individuals to occasionally adopt strategies with lower payoffs, reflecting bounded rationality and noise in decision-making. This imitation rule is widely used in evolutionary game theory and provides a smooth interpolation between deterministic imitation and random strategy adoption. An example of the strategy update is shown in part b) of Fig.~\ref{FIG:0}.

\subsection{Structural evolution of network}

Based on the payoff $\Pi_i$ obtained in current round, each individual determines the maximum number of public goods groups it can participate in. This is mapped through the function
\begin{equation}
    G_i^{max}=max(\frac{a\Pi_i}{c},1)\,,
\end{equation}
\textcolor{black}{where $a$ is a dimensionless sensitivity parameter that controls how strongly an individual's accumulated payoff affects its group-participation capacity. It does not correspond to a directly measurable real-world quantity, but rather serves as a coarse-grained proxy for the responsiveness of resource allocation or organizational expansion. Larger values of $a$ indicate a stronger reaction to payoff differences, meaning that successful individuals expand their participation more aggressively, whereas smaller values of $a$ represent a more conservative adjustment. In the simulations, we consider representative values of $a$ spanning weak, moderate, and strong sensitivity regimes.} $c$ denotes the investment required by a cooperator in each group, the division by $c$ reflects that, for a given payoff, the number of groups an individual can join is inversely related to the per-group contribution requirement: higher investment demands limit participation, while lower demands allow engagement in more groups. This constraint applies to both cooperators and defectors. Although defectors do not contribute, they cannot freely join an arbitrary number of groups. In reality, organizations typically do not admit members indiscriminately; even if they cannot control whether an individual invests, they selectively admit only those who have the potential or capacity to contribute. Accordingly, defectors must also have the investment capability to join more public goods groups. Since defectors do not actually invest, they tend to accumulate higher payoffs compared with cooperators under the same conditions and are therefore more likely to join additional hyperedges. The lower bound of $1$ ensures that every individual participates in at least one group, preventing trivial isolation and preserving the fundamental dynamics of the coevolving network. Given the maximum number of groups $G_i^{max}$, individuals will adjust their participation in hyperedges.

Exit Rule: If the number of groups an individual $i$ currently participates in, $G_i$, exceeds $G_i^{max}$, it must leave some groups. The probability of leaving a particular group $g$ is proportional to the difference between the maximum payoff among $i$'s groups and the payoff obtained from the group $g$:
\begin{equation}
    p_{exit}(g)=\frac{\Pi_i^{max}-\Pi_g+\epsilon}{\Sigma_{h \in G_i}(\Pi_i^{max}-\Pi_h+\epsilon)}\,,
\end{equation}
where $\Pi_i^{max}=max_{h \in G_i}\Pi_h$ is the highest group payoff for individual $i$ and $G_i$ is the set of groups in which $i$ participates. Here $\epsilon>0$ is a small smoothing factor to avoid division by zero and at the same time it gives individuals a relatively small chance to leave the group that offers the highest benefit.

Joining Rule: If $G_i<G_i^{max}$, the individual can join additional groups. The probability of joining a group $g$ is proportional to the group’s payoff in the previous round:
\begin{equation}
    p_{join}(g)=\frac{\Pi_g+\epsilon}{\Sigma_{h \notin G_i}(\Pi_h+\epsilon)}\,,
\end{equation}
where $\epsilon>0$ ensures that groups with zero payoff still have a nonzero probability of being selected. Individuals preferentially join higher-payoff groups, reflecting realistic behavior where more successful projects attract more participants~\cite{szolnoki_amc20}.

In addition to individual-level adaptation, we further introduce a reconfiguration mechanism for hyperedges to capture the dynamic emergence and collapse of collective activities. Specifically, if a hyperedge contains no cooperators for one or multiple consecutive rounds, the public goods game conducted within this group yields zero payoff. Such hyperedges are regarded as non-productive and are removed from the hypergraph. This mechanism mimics real-world collective projects that collapse due to the absence of cooperation or organization.

For each removed hyperedge, a new hyperedge of the same order is generated to preserve the overall size distribution of hyperedges. The members of the new hyperedge are chosen from the population according to a roulette-wheel selection process, with probabilities proportional to their individual payoffs. Specifically, the probability that individual $i$ is selected as a member of the newly generated hyperedge is given by
\begin{equation}
    p(i)=\frac{\Pi_i-\Pi_{min}+\epsilon}{\Sigma_{j \in N}(\Pi_j-\Pi_{min}+\epsilon)}\,,
\end{equation}
where $\Pi_i$ denotes the payoff of individual and $\Pi_{min}$ is the minimum payoff value in the population. $N$ is the set of all nodes in the hypergraph and $\epsilon>0$ is a small smoothing constant that ensures numerical stability. Newly generated hyperedges will not include duplicate individuals when selecting their members.

This mechanism reflects the realistic tendency that individuals with higher accumulated resources are more capable of initiating or being involved in new collective activities.

Notably, a hyperedge is allowed to persist even if it contains only a single individual. In this case, the hyperedge represents a self-sustained activity without interaction, which may later attract additional participants through the joining dynamics described above. Hyperedges containing only defectors, on the other hand, naturally tend to collapse due to the absence of cooperation-induced benefits. \textcolor{black}{More precisely, if a hyperedge remains devoid of cooperators for $\tau$ consecutive rounds, it is removed from the hypergraph. Here, $\tau$ is a time threshold that measures the persistence of an unproductive hyperedge before collapse; a smaller $\tau$ implies a less tolerant and more fragile group structure, whereas a larger $\tau$ indicates stronger inertia or resilience.} Illustrative examples of network structural adaptation and hyperedge reconfiguration are shown in part~c) and d) of Fig~\ref{FIG:0}.

\subsection{Integration of coevolution process}

At each discrete time step, the evolutionary dynamics proceeds as follows:
\begin{itemize}
\item[1)] Public goods game. Given the current hypergraph structure and strategies, public goods games are played independently on all hyperedges. Each individual obtains its payoff $\Pi_i$ by accumulating the contributions from all groups it participates in.

\item[2)] Strategy update. Individuals update their strategies via an imitation, by comparing their payoff with that of a randomly selected neighbor sharing at least one common hyperedge.

\item[3)] Payoff–capacity mapping. Based on the payoff obtained from the public goods game, each individual calculates the maximum number of groups it can participate in, denoted by $G_i^{max}$.

\item[4)] Network adaptation. Individuals compare their current number of groups $G_i$ with $G_i^{max}$. If $G_i>G_i^{max}$, individuals leave the present groups according to the exit rule biased toward low-payoff groups. If $G_i<G_{max}$, individuals join new groups following the payoff-based preferential attachment rule.

\item[5)] Hyperedge reconfiguration process. Hyperedges that contain no cooperators for a prescribed period $\tau$ are removed. For each removed hyperedge, a new hyperedge of the same order is created, whose members are selected via payoff-proportional roulette-wheel sampling from the population.

\end{itemize}

\section{Simulations}

In this section, we first present the methodology employed for the simulation of the coevolutionary dynamics, along with the initial network configuration and parameter settings. We then illustrate the temporal evolution of the cooperation level and the average node degree during the coevolutionary process. Particular attention is devoted to analyzing their steady-state values after the system reaches equilibrium, as well as their time-varying characteristics throughout the evolution, and to how these properties are influenced by the key parameters of the proposed model.

\subsection{Methods}

In the Monte Carlo simulations the initial network consists of 1000 nodes and 300 hyperedges where each hyperedge connecting three nodes selected uniformly at random. Each node is assigned as a cooperator or a defector with equal probability. In the imitation rule, $\beta$ is set to 1; the time threshold $\tau$ for the reorganization of all-defector hyperedges is set to 1000 time steps; the smoothing parameter $\epsilon$ is set to 0.01; and the contribution cost of cooperators is fixed at $c=1$.

The network evolves in discrete time steps under an asynchronous updating scheme. At each time step, the group payoff within each hyperedge is first calculated, followed by the computation of individual payoffs for all nodes. Subsequently, a node is randomly selected to update its strategy according to the imitation rule. Then, another node is randomly chosen, and its maximum allowable number of hyperedges is determined based on its current state, after which its hyperedge connections are updated accordingly. Finally, all hyperedges in the network are examined: those that contain only defectors for an extended period are removed and their member nodes are reassigned to newly formed hyperedges.

\subsection{Evolution of cooperation and average degree over time}

\begin{figure}[]
	\centering
		\includegraphics[scale=.47]{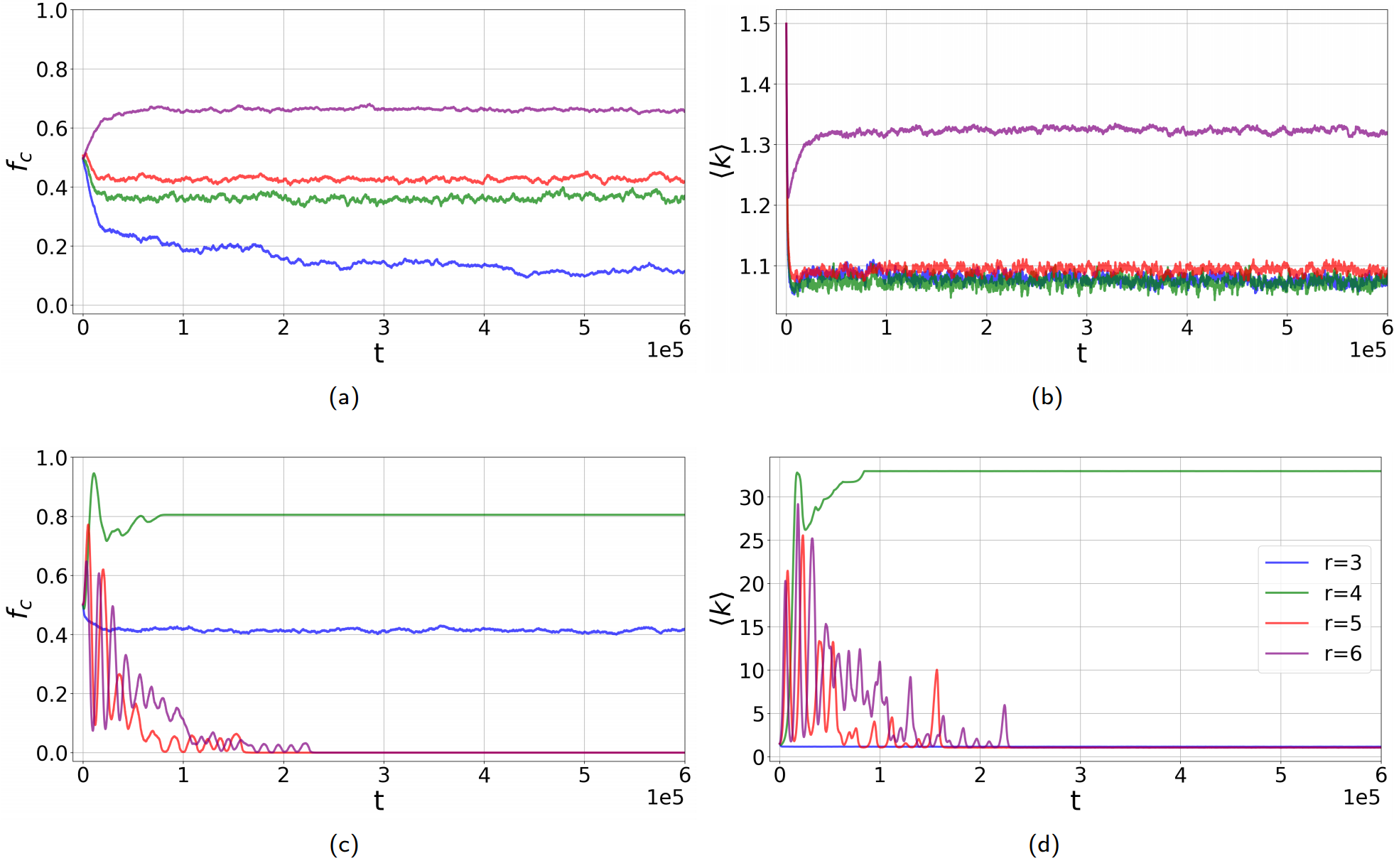}
	\caption{\textbf{Evolution of the cooperation fraction and the average hyperdegree.} Panels~(a) and (b) correspond to the case $a=0.2$, while panels~(c) and (d) correspond to $a=0.5$. Panels~(a) and (c) show the temporal evolution of the cooperation fraction, whereas panels~(b) and (d) display the evolution of the average hyperdegree. In each panel, the blue, green, red, and purple curves correspond to $r=3$, $r=4$, $r=5$, and $r=6$, respectively.}
	\label{FIG:1}
\end{figure}

\begin{figure}[]
	\centering
		\includegraphics[scale=.37]{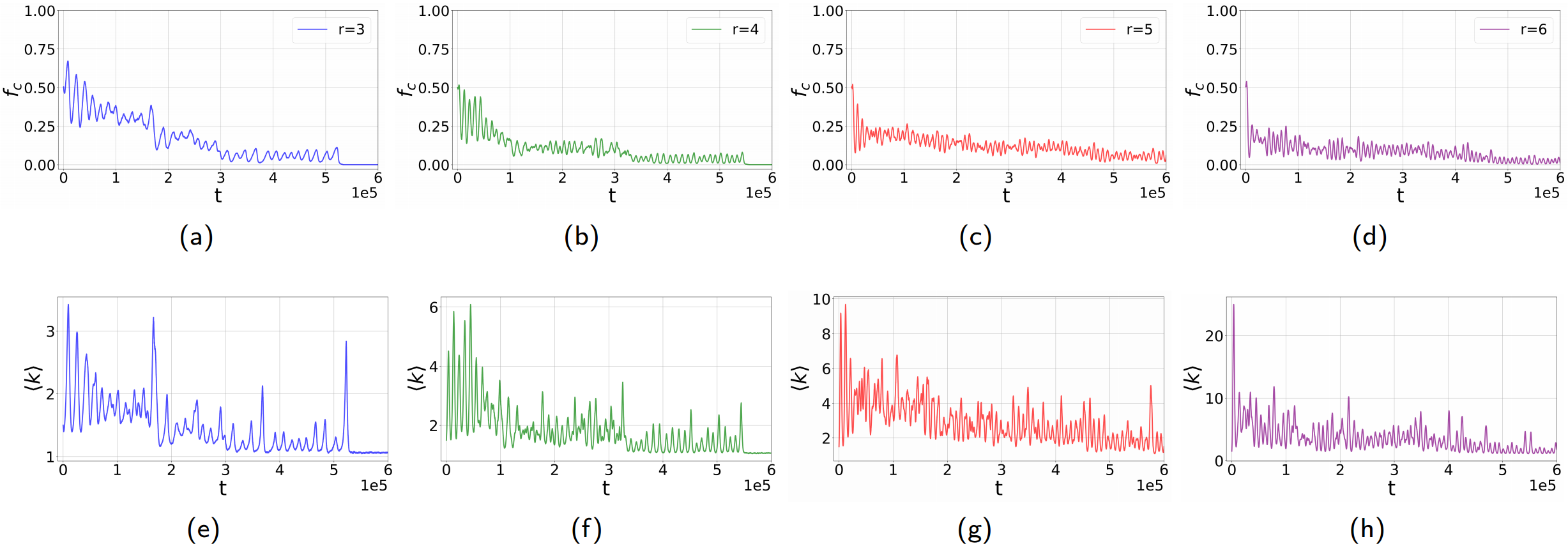}
	\caption{\textbf{Evolution of the cooperation fraction and the average hyperdegree for $a=0.8$.} Panels~(a)-(d) show the evolution of the cooperation fraction, while panels~(e)-(h) display the evolution of the average hyperdegree. In each panel, the blue, green, red, and purple curves correspond to $r=3$, $r=4$, $r=5$, and $r=6$, respectively.}
	\label{FIG:2}
\end{figure}

In this subsection, we present the temporal evolution of the fraction of cooperators, denoted by $f_c$, and the average hyperdegree of nodes, denoted by $\langle k \rangle$. Fig.~\ref{FIG:1} illustrates the evolutionary dynamics for $a=0.2$ and $a=0.5$ under different values of $r$, while Fig.~\ref{FIG:2} shows the corresponding results for $a=0.8$. All results are averaged over more than 10 independent realizations and each simulation is run for $6\times10^5$ time steps.

As shown in Figs.~\ref{FIG:1} and \ref{FIG:2}, both the cooperation fraction and the average hyperdegree converge to stable states after sufficiently long evolution, namely after $6\times10^5$ time steps. From Fig.~\ref{FIG:1}(a) and Fig.~\ref{FIG:1}(b), for the case $a=0.2$, a larger value of $r$ leads to a higher steady-state cooperation fraction. Meanwhile, the average hyperdegree decreases rapidly at the early stage of the evolution and then remains at a very low level. Specifically, for $r=3$, $r=4$, and $r=5$, the average hyperdegree fluctuates around 1, while for $r=0.6$, it is slightly higher and can reach approximately 1.3. This extremely sparse network structure is mainly due to the small value of $a$. A smaller $a$ means that individuals require a much higher payoff to establish new hyperedges. As a result, most nodes fail to get sufficient payoff, either because the initial number of hyperedges is too small or because interactions with defectors reduce their payoff, so they cannot create additional connections. Consequently, the average hyperdegree continuously decays during the evolution.

From Fig.~\ref{FIG:1}(c) and Fig.~\ref{FIG:1}(d), when $a=0.5$, a larger value of $r$ may instead lead to the extinction of cooperators. When $r=3$, cooperation can be maintained at a level of approximately $f_c\approx0.4$, whereas when $r=4$, a much higher cooperation level is observed. In the case where cooperation disappears, the average hyperdegree is also very low. Only for $r=4$ does the network exhibit a relatively high average hyperdegree, exceeding 30, and the evolution becomes frozen after a transient period. This indicates that, under such conditions, hyperedges tend to concentrate among cooperators, forming a dense and prosperous cooperative cluster, while leaving many defectors isolated. Due to the large payoff difference, defectors can hardly invade the cooperative cluster again, which leads to the freezing of the evolution. It is also worth noting that for $r=3$, although cooperation persists, the average hyperdegree remains low. This suggests that cooperation is preserved not because of a well-connected cooperative structure, but because the network structure collapses, isolating both cooperators and defectors, and thereby protecting cooperators.

In addition, for sufficiently large values of $r$, the system exhibits pronounced fluctuations during evolution, and both $f_c$ and $\langle k \rangle$ eventually approach zero. This is because an excessively large $r$ allows individuals to accumulate high payoff values very quickly, which accelerates the formation of new connections and the expansion of hyperedges. Consequently, the interaction probability between cooperators and defectors increases substantially, ultimately making cooperation unsustainable.

For the case $a=0.8$, as shown in Fig.~\ref{FIG:2}, cooperation cannot be sustained regardless of the value of $r$, and the system exhibits pronounced quasi-periodic oscillations throughout the evolution. The failure of cooperation is caused by a mechanism similar to that observed for large $r$ when $a=0.5$: a larger $a$ also accelerates link formation, thus increasing the probability of interactions between cooperators and defectors and destabilizing cooperation. Moreover, the long-term quasi-periodic oscillations suggest that the system may evolve around a specific region in the state space, possibly forming a limit cycle. This issue will be further investigated in the next section.

\subsection{Trajectory of the network in phase space}

\begin{figure}[]
	\centering
		\includegraphics[scale=.39]{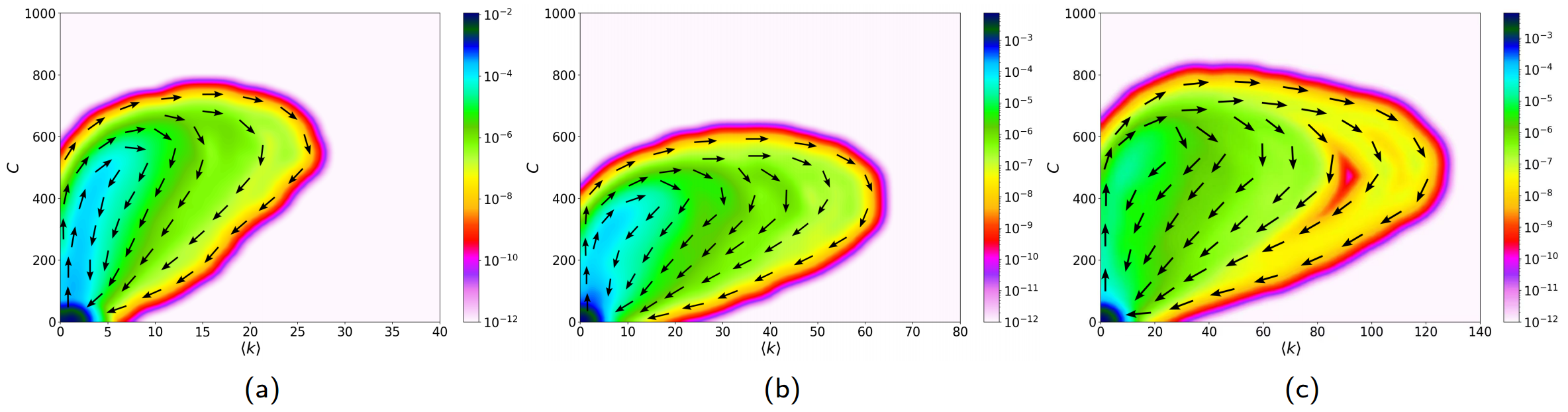}
	\caption{\textbf{Phase-space visitation density of the hypergraph during the evolution.} The horizontal axis represents the average hyperdegree, and the vertical axis denotes the number of cooperators; together they define the phase space of the system. Panels~(a), (b), and (c) correspond to the cases $r=4, a=0.8$; $r=6, a=0.8$; and $r=6, a=0.6$, respectively. The color scale from white to red, green, and blue indicates increasing visitation density. The black arrows mark the direction of state transitions along the evolutionary trajectories.}
	\label{FIG:3}
\end{figure}

In this subsection, we investigate the periodic oscillatory behavior of the system. The probability of visiting a state induced by coevolutionary dynamics is illustrated in Fig.~\ref{FIG:3}. Specifically, Fig.~\ref{FIG:3} presents the phase-space visitation probability of the network, where the vertical axis denotes the number of cooperators, and the horizontal axis represents the average hyperdegree of nodes. These two variables jointly define the phase space of the system, characterizing the possible states visited during the evolutionary process. All results are obtained by aggregating data from ten independent simulation runs. Since all simulations share the same initial condition, the first 300 time steps are discarded to eliminate transient effects, and only the subsequent evolution is used for statistical analysis.

From Fig.~\ref{FIG:3}, it can be seen that, under all parameter settings, the system exhibits petal-shaped limit cycles in the phase space, with trajectories evolving in a clockwise direction. Starting from a state with a medium number of cooperators, individuals tend to obtain higher payoffs, which promotes the expansion of their connections and leads to an increase in the average hyperdegree. As the network becomes denser, the interaction probability between cooperators and defectors increases, resulting in a gradual decline in the number of cooperators. Consequently, individual payoffs decrease, making it difficult to sustain a high number of hyperedges, and the average hyperdegree correspondingly decreases. As the hyperdegree decreases, the network becomes sparse, leading to the formation of small cooperative clusters in which cooperators are effectively protected. Within these clusters, cooperators are able to maintain a certain level of payoff, which enables them to gradually expand their connections again, join new hyperedges and spreading cooperative behavior. As a result, the overall cooperation level of the network increases. The system repeatedly undergoes this cyclic process, thereby forming a stable limit cycle in the phase space. It should be noted that, due to the stochasticity in both strategy updating and structural evolution, the system may occasionally reach absorbing states. Specifically, cooperators may become extinct during the declining phase, driving the system into a fully defective absorbing state.

Furthermore, by comparing the phase-space visitation probability under different parameter settings in Fig.~\ref{FIG:3}, it can be found that when $a=0.8$, increasing $r$ from 4 to 6 leads to a higher maximum average hyperdegree attained during evolution. In addition, when $r=6$, decreasing $a$ from 0.8 to 0.6 also increases the maximum average hyperdegree. These observations indicate that the oscillatory behavior of the system does not vary monotonically with respect to $a$, but is jointly determined by both $a$ and $r$. The amplitude of oscillations is governed by how easily individuals can acquire and sustain sufficient payoff within the network.

\subsection{Cooperation fraction and network density in the parameter space}

\begin{figure}[]
	\centering
		\subfloat[]{\includegraphics[scale=.49]{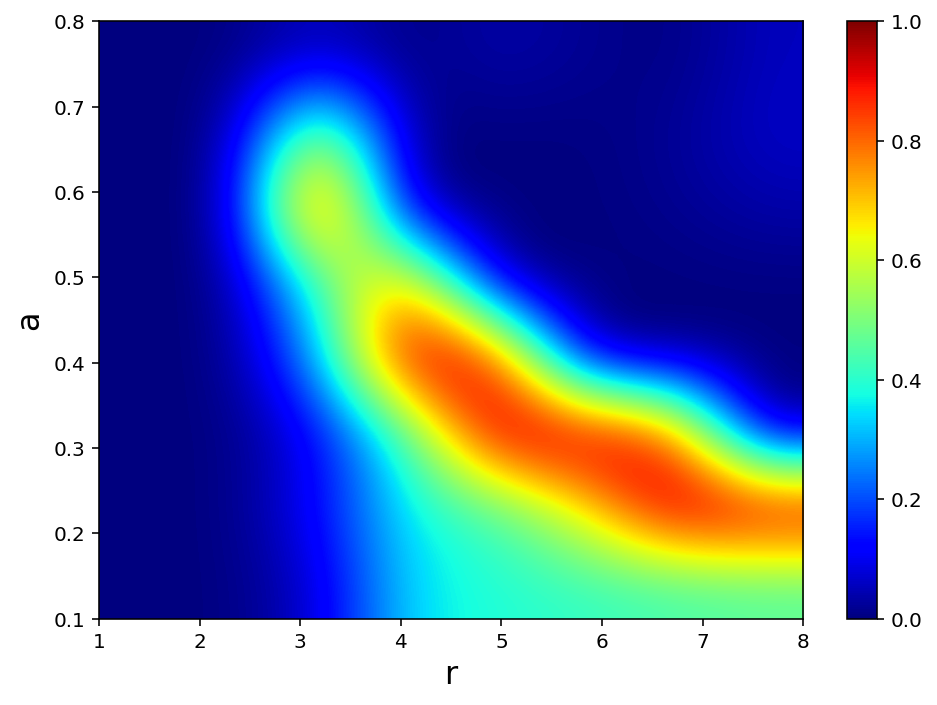}}
        \subfloat[]{\includegraphics[scale=.49]{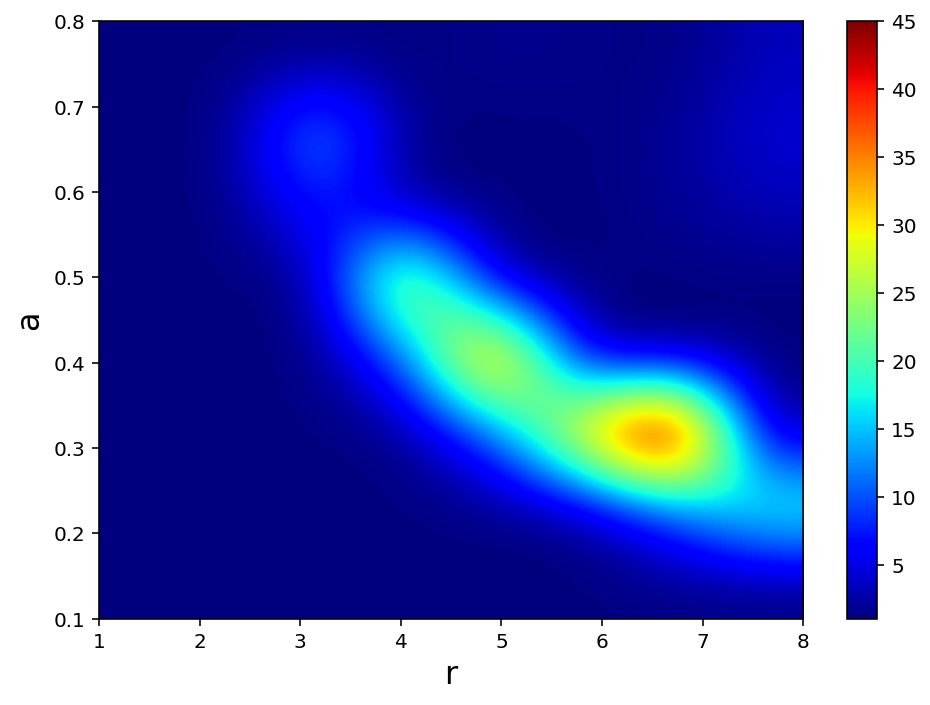}}
	\caption{\textbf{Heat maps of the cooperation fraction and the average hyperdegree in the parameter space.} Panel~(a) corresponds to the cooperation fraction, while panel~(b) corresponds to the average hyperdegree. The horizontal axis represents $r$, and the vertical axis represents $a$, which together define the parameter plane of the model. The color scale from blue to green and red indicates increasing values of the steady-state cooperation level or average hyperdegree under the corresponding parameter combinations.}
	\label{FIG:4}
\end{figure}

In this subsection, we present the heat maps of the steady-state cooperation level and the average hyperdegree of the network in the parameter space, as shown in Fig.~\ref{FIG:4}. As previously noted, each pixel in Fig.~\ref{FIG:4} corresponds to the average over more than 10 independent simulation runs, and each run lasts for more than $6\times10^5$ time steps. Since the network exhibits oscillatory behavior, we first compute the time average of the cooperation fraction and the average hyperdegree over the last 5000 steps of each run, and then average these values over all independent simulations to obtain the presented values.

For the cases in which the hypergraph reaches the fully cooperative state, the hyperedges may continue to expand indefinitely over time until they eventually include all nodes in the hypergraph, which leads to excessively large values of the average hyperdegree. Such results are not meaningful for presentation and may obscure other patterns in the figure. Therefore, in this case, we truncate the evolution once the network reaches the fully cooperative absorbing state, and directly sample the cooperation level and average hyperdegree at that moment.

As can be seen in Fig.~\ref{FIG:4}, the effects of parameter~$a$ and $r$ on both the cooperation fraction and the average hyperdegree are highly nonlinear. Relatively high cooperation levels are concentrated around an arc-shaped band in the parameter space, and the distribution of the average hyperdegree is similar, although its peak values are more localized. This behavior arises because both $a$ and $r$ affect the expansion of individual connections during evolution. When both parameters are large, the frequent formation of new links causes extensive interactions between cooperators and defectors, which eventually destroys cooperation, lowers individual payoffs, and prevents the network structure from being sustained. When $r$ is too small, the payoff of cooperators is insufficient and cooperation cannot be maintained. Only when the combined effects of $a$ and $r$ keep the formation of new links at an appropriate rate can cooperation persist and develop on the network, thus also promoting a denser network structure.

It is worth noting that when $a$ is very small while $r$ is relatively large, the network may still maintain a certain cooperation fraction despite having a very low average hyperdegree. This is because a small $a$ implies that individuals require a sufficiently high payoff to expand and maintain their connections. As a result, the network structure collapses rapidly, a large number of nodes become isolated, and the evolution becomes effectively frozen, thereby protecting the cooperators.

\subsection{Oscillations of cooperation fraction and network density in parameter space}

Motivated by the quasi-periodic oscillations observed above, we further quantify the regularity of the network dynamics by introducing the spectral entropy (SE)~\cite{shannon1948mathematical,bein2006entropy,powell1979spectral}. Let $x(t)$ denote the time series of an observable of interest, namely $f_c(t)$ or $\langle k \rangle(t)$, extracted from the coevolutionary process. After removing the mean value, we compute the discrete Fourier transform (DFT) of the time series $x(t)$,
\begin{equation}
	X(\omega_m)=\sum_{t=0}^{T-1} x(t)\, e^{-i \omega_m t}, \qquad m=0,1,\dots,M-1
\end{equation}
where $T$ is the length of the time series, $M$ is the number of discrete frequency components (frequency bins) used in the entropy calculation, $i=\sqrt{-1}$ is the imaginary unit, and $\omega_m=2\pi m/T$ denotes the discrete angular frequency. Here, $m$ indexes the discrete frequency components (frequency bins). The DFT measures the contribution of each frequency component to the original time series by projecting $x(t)$ onto a set of orthogonal complex exponential bases. The corresponding power spectrum is defined as
\begin{equation}
	P(\omega_m)=|X(\omega_m)|^2\,,
\end{equation}
which quantifies the energy associated with each frequency component. We then normalize the power spectrum to obtain a probability distribution
\begin{equation}
p_m=\frac{P(\omega_m)}{\sum_{j=0}^{M-1}P(\omega_j)}, \qquad m=0,1,\dots,M-1
\end{equation}
The spectral entropy is defined as
\begin{equation}
S_{\mathrm{SE}}=-\frac{1}{\ln M}\sum_{m=0}^{M-1} p_m\ln p_m\,,
\end{equation}
which satisfies $0\le S_{\mathrm{SE}}\le 1$. A smaller value of $S_{\mathrm{SE}}$ indicates that the spectral energy is concentrated in a few dominant frequencies, corresponding to more regular and periodic oscillations, while a larger value implies a broader spectral distribution and therefore more irregular and noisy fluctuations. This makes SE particularly suitable for our system, a frequency-domain entropy provides a compact scalar measure to distinguish whether the system exhibits random fluctuations, as observed in Figs.~\ref{FIG:1}(a) and (b), or develops a limit cycle with periodic oscillations, as shown in Figs.~\ref{FIG:2} and \ref{FIG:3}. Accordingly, for each parameter setting in the phase diagram, we compute $S_{\mathrm{SE}}$ from the last 5000 time steps of each run; when the dynamics reaches an absorbing state, we instead use the 5000 time steps immediately preceding the freezing event. 

\begin{figure}[]
	\centering
		\subfloat[]{\includegraphics[scale=.49]{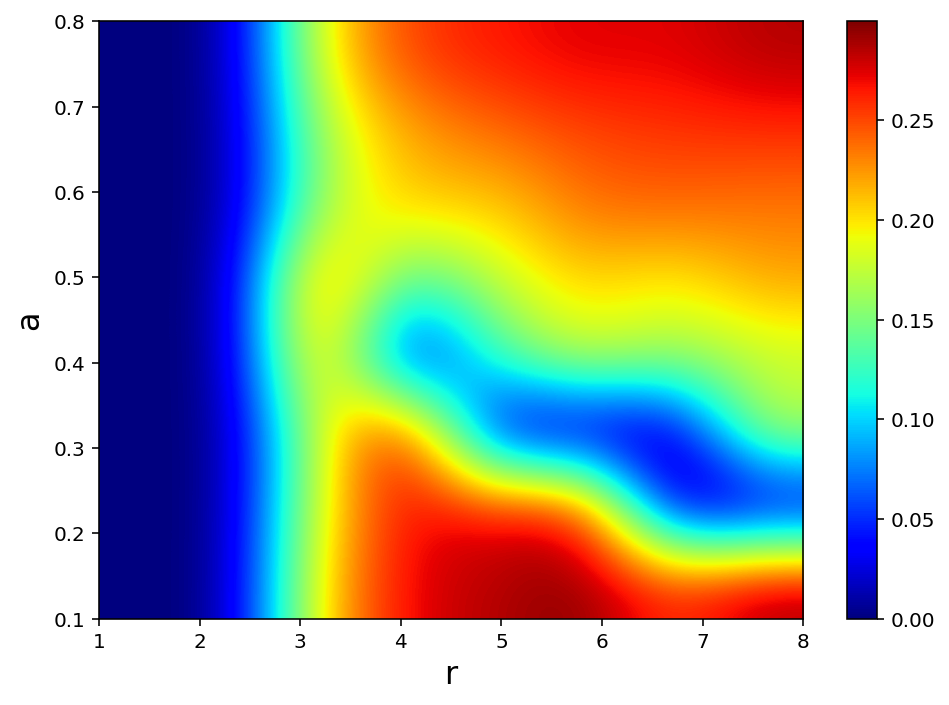}}
        \subfloat[]{\includegraphics[scale=.49]{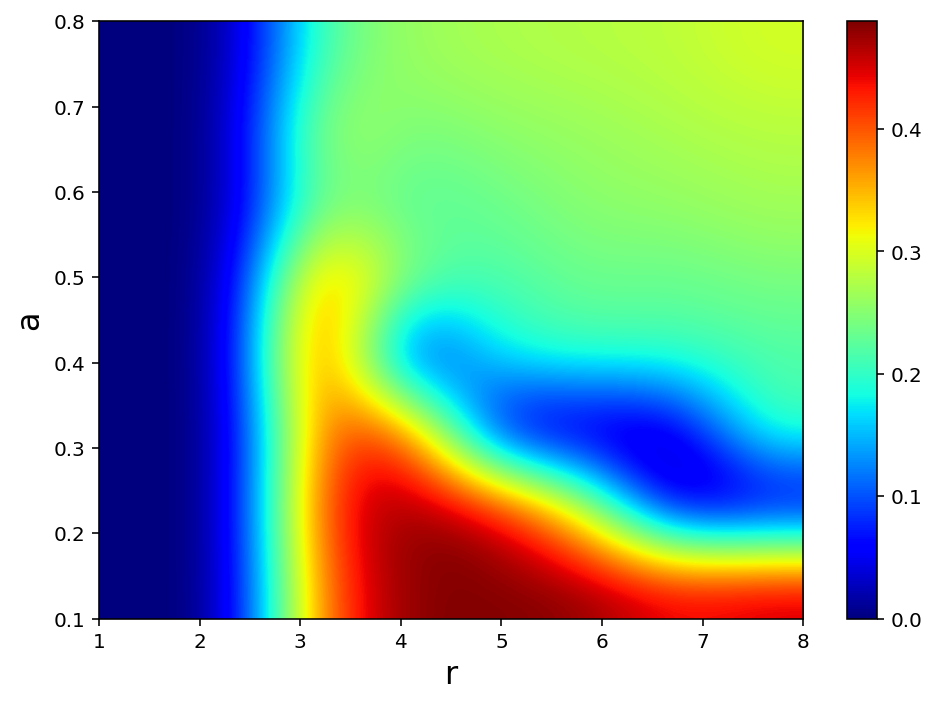}}
	\caption{\textbf{Spectral entropy of the cooperation level and the average hyperdegree in the parameter space.} Panel~(a) corresponds to the cooperation fraction, while panel~(b) corresponds to the average hyperdegree. The horizontal axis represents $r$, and the vertical axis represents $a$, which together define the parameter plane of the model. The color scale from blue to green and red indicates increasing spectral entropy values of the cooperation level or average hyperdegree, computed from the last 5000 time steps of the evolution, or from the 5000 time steps preceding the freezing point when the system reaches an absorbing state.}
	\label{FIG:5}
\end{figure}

The corresponding results are presented in Fig.~\ref{FIG:5}, where each pixel represents the average over more than 10 independent simulation runs. It can be seen that for very small values of $r$, or for cases in which the cooperation fraction is very high, the spectral entropy $S_{\mathrm{SE}}$ of both the cooperation fraction and the average hyperdegree is very low. This is because the system rapidly evolves into either a fully cooperative or a fully defective absorbing state, leading to a quick freezing of the dynamics and, consequently, a low spectral entropy. For the regime with relatively large $r$ and small $a$, the system consistently exhibits high spectral entropy values. This indicates that the dynamics are dominated by irregular and noisy fluctuations rather than periodic behavior, which is consistent with the patterns observed in Figs.~\ref{FIG:1}(a) and (b).

For the cooperation level, low spectral entropy values are mainly concentrated in the intermediate region between the fully cooperative and fully defective states. This suggests that when cooperators and defectors coexist, the system is more likely to exhibit regular, structured oscillations. When both $r$ and $a$ are large, the spectral entropy becomes moderately higher, although it generally remains below 0.3. For most of the parameter combinations, this implies that the cooperation fraction evolves predominantly in an oscillatory manner. For the average hyperdegree, when $a<0.2$ and $r>2$, the spectral entropy is typically above 0.4, indicating more irregular fluctuations. When both $r$ and $a$ are large, the spectral entropy is generally around 0.25, which is comparable to that of the cooperation fraction. This suggests that, in this regime, the evolution of the average hyperdegree is also close to periodic oscillations.

\section{Conclusion and Outlook}

In this paper, we proposed a coevolutionary public goods game model on a dynamic hypergraph, in which an individual's payoff determines the number of hyperedges that it can participate in. In other words, the payoff, as a direct marker of success, controls how many group interactions the individual is able to join. When a node enters or leaves a hyperedge, it makes its decision based on the group payoff of that hyperedge. Moreover, if a hyperedge remains occupied only by defectors for a sufficiently long time, it collapses and is reorganized by selecting a new set of members according to the current payoffs of nodes in the network. This mechanism reflects a common reorganization process in real group-structured systems, where persistently unproductive groups tend to dissolve and be rebuilt through performance-based selection. This mechanism is consistent with many practical scenarios, including temporary research collaborations, project-based teams in organizations, online communities with fluctuating participation, and cooperative partnerships in which membership is continually adjusted according to observed contribution or benefit.

Within this framework, we investigated the temporal evolution of the cooperation fraction and the average hyperdegree. The results show that the collective dynamics can be either irregularly fluctuating or periodically oscillatory, depending on the parameter regime. In particular, cooperation persists only when the parameters $r$ and $a$ jointly produce a suitable rate of link formation. If structural adaptation is too fast, cooperators and defectors interact too frequently, which destroys cooperative clusters and suppresses cooperation. If the adaptation is too slow, the cooperators cannot maintain sufficient structural support to expand. Therefore, cooperation is promoted not by maximizing the speed of structural growth, but by keeping it within an appropriate range. This nonmonotonic dependence differs from the conventional conclusion in static evolutionary game models, where increasing the benefit parameter usually facilitates cooperation more directly. In the proposed model, an excessively large benefit parameter may lead to overly rapid mixing among individuals, thus exerting a negative effect on the emergence and maintenance of cooperation. This observation is relevant to realistic systems in which overly frequent interaction or overly rapid reorganization may reduce, rather than enhance, collective efficiency, such as team collaboration under excessive member turnover, volunteer groups with unstable participation, or social networks with rapid exposure to low-quality interaction partners.

To characterize the oscillatory behavior of the system, we further introduced the spectral entropy. The numerical results indicate that when both $r$ and $a$ are large, the dynamics tend to form a limit cycle in the phase space, producing a repeated sequence of cooperation growth, network densification, cooperation suppression, network sparsification, and cooperation recovery. Such recurrent behavior is reminiscent of cyclic adaptation observed in many real systems, where successful cooperation promotes expansion, while excessive expansion increases conflict, competition, or overload, eventually weakening cooperation and forcing the system to reorganize. Examples include periodic formation and dissolution of collaborative clusters in social and organizational settings, repeated congestion-relief cycles in resource-sharing systems, and adaptive restructuring in collective decision-making environments. From this perspective, the model highlights how adaptive restructuring can simultaneously support cooperation and generate instability and provides an insightful explanation for why oscillatory cooperation may persist in group-based interaction systems.

Several extensions are worth pursuing in future work. \textcolor{black}{First, the present payoff-to-capacity mapping is piecewise linear, $G_i^{max}=max(a\Pi_i/c,1)$, which may be too abrupt for some real-world systems. A natural extension is to replace it with a smoother sigmoidal mapping, such as a logistic or tanh-type function. It would be interesting to examine whether such a smooth response weakens the oscillatory behavior observed in the present model or shifts the phase boundaries between oscillatory and non-oscillatory regimes. Second, introducing heterogeneity may reveal richer collective outcomes. For example, one may assume that the payoff sensitivity $a_i$ follows a distribution rather than taking a homogeneous value, and then test whether this heterogeneity leads to polarization in cooperation levels, with highly responsive individuals rapidly expanding their participation while less responsive individuals remain trapped in low-participation states. Third, memory effects provide another promising direction. In the present model, individuals react to their current payoff, but in realistic settings decisions often depend on historical experience. Replacing the instantaneous payoff by a moving average of past payoffs may smooth the response to fluctuations, suppress rapid switching of group participation, and possibly weaken or delay oscillations. Finally, the hyperedge reorganization rule can be further refined by incorporating preference, spatial constraints, or empirical formation data, while a deeper theoretical analysis of cooperation thresholds, stability conditions, and absorption times remains an important open problem.}

\section{Acknowledgement}
The authors thank the support from the National Natural Science Foundation of China under grant Nos. 12272282, and Natural Science Foundation of Chongqing under Grant CSTB2025YITP-QCRCX0007, and National Research, Development and Innovation Office (NKFIH) under Grant No. K142948.

\bibliographystyle{model1-num-names}


\end{document}